\documentclass[showpacs,aps]{revtex4}
\usepackage{epsfig}

\newcommand{\be}{\begin{equation}}
\newcommand{\ee}{\end{equation}}
\newcommand{\ba}{\begin{eqnarray}}
\newcommand{\ea}{\end{eqnarray}}

\newcommand{\Li}{{\rm Li}}

\newcommand{\la}{\left\langle}

\newcommand{\dd}{\mathrm d}

\def\DD{{\mathcal D}}
\def\SS{{\mathrm{S}}}
\def\NS{{\mathrm{NS}}}

\def\MSbar{$\overline{\mathrm{MS}}\ $}
\def\Sot#1{{\mathrm{S}}_{1,2}\left(#1\right)}
\def\order#1{{\mathcal{O}}\left(#1\right)}
\def\ba{\begin{eqnarray}}
\def\ea{\end{eqnarray}}

\def\la{\mathrel{\mathpalette\fun <}}

\def\fun#1#2{\lower3.6pt\vbox{\baselineskip0pt\lineskip.9pt
  \ialign{$\mathsurround=0pt#1\hfil##\hfil$\crcr#2\crcr\sim\crcr}}}
\def\order#1{{\mathcal O}\left(#1\right)}

\begin{document}

\preprint{SLAC-PUB-9220, Alberta Thy 09-02, hep-ph/0205172}

\title{
\[ \vspace{-2cm} \]
\noindent\hfill\hbox{\rm Alberta Thy 09-02} \vskip 1pt
\noindent\hfill\hbox{\rm SLAC-PUB-9220} \vskip 1pt
\noindent\hfill\hbox{\rm hep-ph/0205172} \vskip 10pt
${\cal O}(\alpha^2 \ln(m_\mu/m_e))$ Corrections to 
Electron Energy Spectrum in Muon Decay}

\author{Andrej Arbuzov}
\email{aarbuzov@phys.ualberta.ca}
\thanks{on leave of absence from JINR, Dubna.}
\affiliation{
Department of Physics, University of Alberta\\
Edmonton, AB\ \  T6G 2J1, Canada}

\author{Kirill Melnikov}
\email{melnikov@slac.stanford.edu}
\affiliation{Stanford Linear Accelerator Center\\
Stanford University, Stanford, CA 94309}

\begin{abstract}
${\cal O}(\alpha^2 \ln(m_\mu/m_e))$ corrections 
to electron energy spectrum in muon decay are computed 
using  perturbative fragmentation function approach.
The magnitude of these corrections is comparable to  
anticipated precision of the TWIST experiment at TRIUMF
where Michel parameters will be extracted from the 
measurement of the electron energy spectrum in muon decay. 
\end{abstract}

\pacs{13.35.Bv, 14.60.Ef, 12.20.Ds}

\keywords{muon decay, radiative corrections}

\maketitle

\section{Introduction}

The muon decay into an electron and a pair of neutrinos, 
$\mu \to e \nu_\mu \bar \nu_e$,  is a classical
process in particle physics. Although the high energy frontier 
has moved up from the energy scales comparable to the  muon mass, 
precision physics of muons remains an interesting and inspiring 
source of information about the Standard Model (SM) and its possible 
extensions~\cite{Kuno:1999jp}.

Among very different experiments with muons 
that include the measurement of   
muon anomalous magnetic moment, muon lifetime, $\mu \to e \gamma$ branching 
ratio, and the muon to electron conversion rate in muonic atoms, 
our attention  will be focused on the TWIST 
experiment~\cite{Rodning:2001js,Quraan:2000vq}
at TRIUMF, where the electron energy spectrum 
in muon decays is planned to be measured to determine Michel parameters
with the precision  of $\sim 10^{-4}$. 
In order to confront these measurements with the SM predictions
and look for the signs of New Physics, 
one needs an adequately accurate calculation of 
the electron energy spectrum within the SM.

Calculations of  the electron energy spectrum in muon decay have a long 
and interesting history that dates back to the very early days 
of QED and the physics of weak interactions 
(see e.g. Ref.~\cite{Kinoshita:2001pn} for historical recollection). 
In spite of tremendous progress in precision calculations, 
the ${\cal O}(\alpha^2)$ radiative corrections to total muon 
lifetime
have been computed only recently \cite{vanRitbergen:1998yd}, 
and the calculation of similar 
corrections to electron energy spectrum has even not been attempted.
One reason for this is that, in contrast to total lifetime, the 
electron energy spectrum in muon decay can not be computed for 
the vanishing  mass of the electron since terms enhanced 
by the large logarithm $\ln(m_\mu/m_e)$ are present. 
These terms (excluding the ones that are related to the on--shell
definition of the fine--structure constant commonly used in QED)
cancel out in the calculation of the total rate which,
for this reason, becomes somewhat simpler. 

At order ${\cal O}(\alpha^2)$,  corrections to the spectrum have 
double--logarithmic ${\cal O}(\ln^2(m_\mu^2/m_e^2))$  
and single--logarithmic ${\cal O}(\ln(m_\mu^2/m_e^2))$
enhanced terms and it is the purpose of this Letter to present 
the calculation of those. The double--logarithmic terms were
computed recently in Ref.~\cite{Arbuzov:2002pp}. It was
pointed out there that the 
single--logarithmic ${\cal O}(\ln(m_\mu^2/m_e^2))$
terms are required to match the precision of 
the   TWIST experiment.
Motivated by these considerations, we decided to perform this 
calculation. In order to accomplish this, we make use of  
perturbative fragmentation function  approach borrowed 
to a large extent from QCD studies of heavy quark 
fragmentation in $e^+e^-$ collisions.

\section{Preliminaries}

According to QCD factorization theorems~\cite{Ellis:qj}, 
which can be transformed to QED in a straightforward way, 
a  differential cross section for producing a particle 
of a given type with a certain fraction of the initial energy 
can be written as a convolution of the hard scattering cross section 
computed  with massless partons and the fragmentation function that 
describes the  probability that a massless parton  of a given type 
fragments to the physical particle in the final state that is being  
observed.

If one considers the process in which a heavy quark (i.e. 
$m_Q \gg \Lambda_{\rm QCD}$) is produced with large energy and 
its energy is measured, one can identify the {\it massive} quark 
with the physical particle in the final state, in the sense of the
preceding  discussion. It has been shown in QCD that in this case 
the perturbative 
fragmentation function can be defined and that this function
absorbs all the terms that are singular in the limit when 
the mass of the heavy quark goes to zero
\cite{Mele:1990cw,Cacciari:2001cw,Corcella:2001hz}. 

Specializing to the case of muon decay, and working along these 
ideas, it is possible to write down the formula for the electron 
energy spectrum in muon decay in the following way:
\ba \label{master}
\frac{\dd \Gamma}{\dd x}(x,m_\mu,m_e) =
\sum\limits_{j=e,\gamma}^{}
\int\limits_{x}^{1}\frac{\dd z}{z}\;
\frac{\dd \hat{\Gamma}_j}{\dd z}(z,m_\mu,\mu_f)
\DD_j\biggl(\frac{x}{z},\mu_f,m_e\biggr),
\ea
where $z =2E/m_\mu$ is the fraction of energy carried away by a 
parton $j$ in muon decay, $x$ is the same quantity for the observed
physical massive electron,   ${\dd \hat{\Gamma}_j}/{\dd z}$ 
is the energy distribution  of the massless parton of type $j$ in muon decay
computed in the \MSbar scheme, $\DD_j$ is the fragmentation 
function of the parton $j$ fragmenting into the massive electron,
and  $\mu_f$ stands for the factorization scale.
Note that the 
terms that are suppressed by the ratio of the electron to muon 
masses,  $m_e^2/m_\mu^2$,  can not be described by 
Eq.~(\ref{master}). However, since  these terms are known, 
both for Born and ${\cal O}(\alpha)$ corrected 
electron energy spectrum~\cite{Kinoshita:1959ru,Arbuzov:2001ui},  
Eq.~(\ref{master}) is quite adequate for anticipated  level of 
experimental precision.

As we mentioned earlier, partonic decay rate ${\dd \hat{\Gamma}_j}/{\dd z}$ 
has to be computed in the $\overline {\rm MS}$ scheme. 
This is important in that this requirement goes beyond 
standard ultraviolet renormalization, since 
${\dd \hat{\Gamma}_j}/{\dd z}$ 
is not finite because of collinear singularities. These 
singularities are removed from ${\dd \hat{\Gamma}_j}/{\dd z}$ 
by conventional  renormalization in the \MSbar  scheme 
and large collinear logarithms associated with them are 
absorbed 
into the fragmentation function $\DD_j$.

The perturbative expansion for the energy distribution of 
the massless partons reads
\ba
&& \frac{1}{\Gamma_0}\frac{\dd\hat{\Gamma}_j}{\dd z }(z,m_\mu,\mu_f)
= A^{(0)}_j(z)
+ \frac{\bar{\alpha}(\mu_f)}{2\pi}\hat{A}^{(1)}_j(m_\mu,\mu_f,z)
+ \biggl(\frac{\bar{\alpha}(\mu_f)}{2\pi}\biggr)^2\hat{A}^{(2)}_j(m_\mu,\mu_f,z),
\ea
where  $\Gamma_0 = G_F^2 m_\mu^5/(96\pi^3)$,  $A^{(0)}_j(z) 
= z^2(3 - 2z)\delta_{je}$, $\bar{\alpha}(\mu_f)$ is the
\MSbar renormalized fine structure constant,
and we have neglected terms of order ${\cal O}(\alpha^3)$ and higher.
The \MSbar fine structure constant will be later  
converted  into the   on--shell coupling fine structure constant 
$\alpha  \approx 1/137.036$.

Before giving explicit expressions for perturbative coefficients, 
we would like to describe  a simple idea (previously
exploited in QCD studies) that allows us to compute the 
$\alpha^2 \ln(m_\mu/m_e)$  enhanced terms without doing explicit two--loop
calculation. 
Since ${\dd \hat{\Gamma}_j}/{\dd z}$  is computed for massless partons, 
it has just two energy  scales, the mass of the muon $m_\mu$ and 
the factorization scale $\mu_f$. This means that
the only logarithms that arise 
there are the logarithms of the form $\ln(m_\mu/\mu_f)$. Therefore, by choosing
$\mu_f \sim m_\mu$, we effectively eliminate large logarithms 
from perturbative coefficients of the energy spectrum  computed with 
massless partons and move all the large logarithms to the fragmentation 
function $\DD_j$. On the other hand, the fact that the fragmentation 
function 
is process--independent and also satisfies the
Dokshitzer--Gribov--Lipatov--Altarelli--Parisi (DGLAP) evolution 
equation  as far as its dependence on the factorization scale is 
concerned, allows us to use some formulas known from previous
QCD studies and  compute large logarithmic corrections 
to the electron energy 
spectrum at order ${\cal O}(\alpha^2)$ 
without performing explicit two--loop calculations.

Let us consider the fragmentation function $\DD_j(x,\mu_f,m_e)$ which 
describes the probability that the massless parton $j$ converts 
into the 
physical  electron of the mass $m_e$. This function satisfies the DGLAP 
evolution equation:
\be
\frac{\dd \DD_i(x,\mu_f,m_e)}{\dd \ln \mu_f^2} = 
\sum \limits_{j} \int \limits_{x}^{1}
\frac{\dd z}{z} P_{ji}(z,\bar{\alpha}(\mu_f) ) 
\DD_j\biggl(\frac{x}{z},\mu_f,m_e\biggr),
\label{DGLAP}
\ee
where $P_{ji}$ stands for  usual time--like splitting functions
which, to the order we work to,  
can be written as
\ba
P_{ji}(x,\bar{\alpha}(\mu_f)) = \frac{\bar{\alpha}(\mu_f)}{2\pi}P_{ji}^{(0)}(x)
+ \biggl(\frac{\bar{\alpha}(\mu_f)}{2\pi}\biggr)^2P^{(1)}_{ji}(x) 
+ \order{\bar \alpha^3}.
\ea
Eq.~(\ref{DGLAP}) can be solved as power series in $\bar \alpha$ if
the initial condition for the function 
$\DD_j$ at some scale $\mu_0$ is provided. 
This initial condition can be obtained  from 
QCD studies of the heavy quark fragmentation~\cite{Mele:1990cw}
which, when translated to QED, imply that the fragmentation 
of the  massless electron into the physical electron is described 
by
\ba \label{FFi}
\DD^{\mathrm{ini}}_e(x,\mu_{0},m_e) &=& \delta(1-x)
+ \frac{\bar{\alpha}(\mu_{0})}{2\pi} d_1(x,\mu_{0},m_e) + \order{\alpha^2}, 
\nonumber \\
d_1(x,\mu_{0},m_e) &\equiv& d_1(x) = \biggl[ \frac{1+x^2}{1-x}
\biggl( \ln\frac{\mu_{0}^2}{m_e^2} - 2\ln(1-x) - 1 \biggr) 
\biggr]_+,
\ea
and  the fragmentation function of the photon fragmenting 
into the physical  electron is 
\ba \label{FFig}
\DD^{\mathrm{ini}}_\gamma(x,\mu_{0},m_e) &=& 
\frac{\bar{\alpha}(\mu_{0})}{2\pi} 
\left ( x^2 + (1-x)^2 \right ) \ln\frac{\mu_{0}^2}{m_e^2}
 + \order{\alpha^2}.
\ea
As will be clear from the following discussion, the ${\cal O}(\alpha^2)$ 
terms in the initial conditions for  fragmentation functions are 
not needed for our purposes since our choice of the
initial scale $\mu_0 \sim m_e$  guarantees that  no 
large logarithms appear in the initial condition for $\DD_j$.
One should also 
notice that at the lowest order ${\cal O}(\alpha^{0})$ the fragmentation 
function does not contain large logarithm and, for this reason,
the second order coefficient  in $\dd \hat \Gamma_j/\dd z$ is not needed
as well.  
On the other hand, the order ${\cal O}(\alpha)$  coefficients in 
 $\dd \hat \Gamma_j/\dd z$ have to be known exactly and they 
read
\ba
\hat{A}^{(1)}_e(z) &=& 
\left ( 2z^2(2z-3)\ln  \left [ \frac{z}{1-z} \right ] 
+ 2z + \frac{8}{3}z^3 + \frac{5}{6} - 4z^2 \right ) 
\ln \left ( \frac{m_\mu^2}{\mu_f^2} \right )
\nonumber \\
&+& 2z^2(2z-3)\left ( 4\zeta_2 - 4\Li_2(z) + 2\ln^2z
- 3\ln z\ln(1-z) - \ln^2(1-z)
\right )
\nonumber \\
&+& \left ( \frac{5}{3} - 2z - 13z^2 
+ \frac{34}{3}z^3 \right )\ln(1-z)
+ \left ( \frac{5}{3} + 4z - 2z^2
- 6z^3 \right )\ln z
\nonumber \\
&+& \frac{5}{6} - \frac{23}{3}z - \frac{3}{2}z^2 + \frac{7}{3}z^3,
\\
\hat{A}^{(1)}_\gamma(z) &=&  
\biggl(\ln\frac{m_\mu^2}{\mu_f^2} + \ln(1-z)\biggr)
\biggl( \frac{1}{z} - \frac{5}{3} + 2z - 2z^2 
+ \frac{2}{3}z^3 \biggr)
+ \ln z \biggl( \frac{2}{z} - \frac{10}{3} + 4z \biggr)
\nonumber \\ 
&-& \frac{1}{z} + \frac{1}{3} + \frac{35}{12}z 
- 2z^2 - \frac{1}{4}z^3.
\ea

With these preliminary remarks out of the way, we can proceed with the 
computation of the fragmentation function and use it to calculate 
the electron energy spectrum in muon decay.

\section{Fragmentation function}

In this Section we discuss the computation of 
the fragmentation 
function. For this purpose, we need to solve the DGLAP 
equation~(\ref{DGLAP})
in a way consistent with the initial conditions.
Since we solve this equation perturbatively, we can express the running 
fine structure  constant in the \MSbar scheme 
through the fine structure constant defined in the on--shell scheme,
because the use of the on--shell renormalization scheme is  
the common practice when dealing with QED corrections.
To order ${\cal O}(\alpha^2)$, the well known relation between 
the \MSbar and the on--shell coupling constants reads
\ba
\bar{\alpha}(\mu_f) = \alpha 
+ \frac{\alpha^2}{3\pi}\ln\frac{\mu_f^2}{m_e^2}\, .
\ea

Solving  Eq.~(\ref{DGLAP}) by iterations, we obtain 
\ba \label{De}
&& \DD_e(x,\mu_f,m_e) = \delta(1-x) 
 + \frac{\alpha}{2\pi} \left ( LP^{(0)}_{ee}(x) 
+ d_1(x,\mu_{0},m_e) \right )
+ \biggl(\frac{\alpha}{2\pi}\biggr)^2
\biggl(
L^2 \biggl[
\frac{1}{2} P^{(0)}_{ee} \otimes P^{(0)}_{ee}(x)
  \nonumber \\ &&  \qquad
+ \frac{1}{3} P^{(0)}_{ee}(x)
+ \frac{1}{2} P^{(0)}_{\gamma e} \otimes P^{(0)}_{e\gamma}(x) \biggr]
+ L \left[ P^{(0)}_{ee} \otimes d_1(x) 
+ P^{(1)}_{ee}(x) \right] \biggr)  + \order{\alpha^2 L^{0}, \alpha^3}, 
\\ \label{Dg}
&& \DD_\gamma(x,\mu_f,m_e) = \frac{\alpha}{2\pi}
LP^{(0)}_{e \gamma}(x) + \order{\alpha^2}, 
\ea
where  $L = \ln(\mu^2_f/\mu_{0}^2)$, and 
the convolution operation is defined in the standard way:
\ba
\label{conv}
A \otimes B(x) = \int\limits^1_0\dd z
\int\limits^1_0\dd z'\; \delta(x-zz')A(z)B(z') 
=\int\limits^1_x\frac{\dd z}{z}\; A(z)B\biggl(\frac{x}{z}\biggr).
\ea

We now give explicit expressions for the splitting functions used 
in Eqs.~(\ref{De},\ref{Dg}).  At leading order they read
\ba \label{P0}
P_{ee}^{(0)}(x) = \biggl[\frac{1+x^2}{1-x}\biggr]_+,
\qquad
P^{(0)}_{\gamma e}(x) = \frac{1 + (1-x)^2}{x}\, ,
\qquad P^{(0)}_{e \gamma}(x) = x^2 + (1-x)^2. 
\ea

At next--to--leading order the time--like splitting functions 
have been derived for QCD in 
Refs.~\cite{Curci:1980uw,Floratos:1981hs,Furmanski:1980cm,
Ellis:1996nn} and, by choosing appropriate 
color structures, can be translated to QED.  Since, experimentally, 
one will probably distinguish events with one and  more electrons 
in the final state, we decided to split the corresponding second 
order function $P^{(1)}_{ee}(x)$ into four parts
in the same way as in Ref.~\cite{Berends:1987ab}:
\ba
P^{(1)}_{ee}(x) = P^{(1,\gamma)}_{ee}(x)
+ P^{(1,\NS)}_{ee}(x)
+ P^{(1,\SS)}_{ee}(x)
+ P^{(1,\mathrm{int})}_{ee}(x).
\ea
Here $P^{(1,\gamma)}_{ee}(x)$ is determined by the set of Feynman
diagrams with pure photonic corrections (i.e. no additional electrons 
in the final state or closed electron  loops
in virtual corrections); $P^{(1,\NS)}_{ee}(x)$ is related to 
corrections due to non--singlet real and virtual $e^+e^-$ pairs;
$P^{(1,\SS)}_{ee}(x)$ stands for the singlet pair production
contribution; and $P^{(1,\mathrm{int})}_{ee}(x)$ describes
the interference of the  singlet and non--singlet pairs.
These functions read
\ba \label{P1g}
P^{(1,\gamma)}_{ee}(x) &=&
\delta(1-x)\biggl( \frac{3}{8} - 3\zeta_2 + 6\zeta_3 \biggr)
+ \frac{1+x^2}{1-x}\biggl( 2\ln x \ln(1-x)
- 2\ln^2x - 2\Li_2(1-x) \biggr)
\nonumber \\ &&
+ \frac{1}{2}(1+x)\ln^2x
+ 2x\ln x - 3x + 2,
\\ \label{P1ns}
P^{(1,\NS)}_{ee}(x) &=& \delta(1-x)\biggl( - \frac{4}{3}\zeta_2 
- \frac{1}{6} \biggr)
- \frac{20}{9}\biggl[\frac{1}{1-x}\biggr]_{+}
- \frac{2}{3} \frac{1+x^2}{1-x}\ln x - \frac{2}{9} + \frac{22}{9}x,
\\ \label{P1s}
P^{(1,\SS)}_{ee}(x) &=& (1+x)\ln^2x
+ \biggl(  - 5 - 9x - \frac{8}{3}x^2 \biggr)\ln x  
- 8 - \frac{20}{9x} + 4x + \frac{56}{9}x^2,
\\ \label{P1int}
P^{(1,\mathrm{int})}_{ee}(x) &=& 
\frac{1+x^2}{1-x}\biggl( 2\Li_2(1-x)
+ \frac{3}{2}\ln x \biggr) - \frac{7}{2}(1+x)\ln x
- 7 + 8x, 
\ea
where we have used
\be
\zeta_n \equiv \sum_{k=1}^{\infty}\frac{1}{k^n}\, ,\qquad 
\zeta_2 = \frac{\pi^2}{6}\, ,\qquad
\Li_2(x) \equiv - \int\limits_{0}^{x}\dd z\;\frac{\ln(1-z)}{z}\, .
\ee

Finally, we give explicit formulas for various convolutions, which 
appear in Eq.~(\ref{De}):
\ba \label{p0p0}
&& P^{(0)}_{ee} \otimes P^{(0)}_{ee}(x)
= \delta(1-x)\biggl( \frac{9}{4} - 4\zeta_2 \biggr)
+ \biggl[\frac{1}{1-x}\biggl( 6 + 8\ln(1-x) \biggr)\biggr]_+
\nonumber \\ && \qquad
- \frac{4}{1-x}\ln x
+ (1+x)[ 3\ln x - 4\ln(1-x) ] - x - 5,
\\ && \label{Rx}
P^{(0)}_{\gamma e} \otimes P^{(0)}_{e\gamma}(x)
= \frac{1-x}{3x}(4+7x+4x^2) + 2(1+x)\ln x,
\\  && \label{p0d1}
P^{(0)}_{ee} \otimes d_1(x) =
P^{(0)}_{ee} \otimes P^{(0)}_{ee}(x)\biggl(
\ln\frac{\mu_{0}^2}{m_e^2} - 1 \biggr)
+ \delta(1-x)\biggl( \frac{21}{4} - 8\zeta_3 \biggr)
\nonumber \\ && \qquad
+ \biggl[ \frac{1}{1-x}\biggl( 7 + 8\zeta_2 
- 6\ln(1-x) - 12\ln^2(1-x) \biggr)\biggr]_+
+ \frac{8}{1-x}\ln x\ln(1-x)
\nonumber \\ && \qquad
+ (1+x)[ 6\ln^2(1-x) - 6\ln x\ln(1-x)
- 2\Li_2(1-x) - 4\zeta_2 ] 
+ 2x\ln x 
\nonumber \\ && \qquad
+ (7-x)\ln(1-x) - \frac{11}{2} - \frac{3}{2}x.
\ea
Using these results in Eqs.(\ref{De},\ref{Dg}), 
one obtains  explicit expressions 
for the fragmentation functions $\DD_{e,\gamma}(x,\mu_f,m_e)$.

\section{Electron energy spectrum}

To obtain  the electron energy spectrum we have to convolute the 
fragmentation functions Eqs.~(\ref{De},\ref{Dg}) with 
$\dd\hat{\Gamma}_j/\dd z$,
and we have given all the necessary ingredients to do that 
in the previous Sections. 
Before presenting corrections to the electron energy spectrum, 
let us note that  the dependence on the 
factorization scale cancels out explicitly in the final result up to 
the terms that are not enhanced by any large logarithm and 
for this reason are beyond the scope of this paper.

As we explained earlier, we split the final result into four different 
pieces: pure photonic radiative corrections, non--singlet pair radiative
corrections, 
singlet pair radiative corrections, 
and corrections due to the interference of singlet 
and non-singlet pairs. Following this separation,  
we write the result for 
the electron energy spectrum in muon decay  
as
\ba
\frac{1}{\Gamma_0} \frac{\dd \Gamma}{\dd x} = 
\Delta^{(\gamma)} + \Delta^{(\NS)} + \Delta^{(\SS)} 
+ \Delta^{(\mathrm{int})}.
\label{final}
\ea

Let us start with  pure photonic radiative corrections. 
Computing the convolutions, 
we arrive at the following result: 
\be
\Delta^{(\gamma)} = f_0(x) + \frac{\alpha}{2\pi}f_1(x) + 
\biggl( \frac{\alpha}{2\pi} \biggr)^2 \left [ 
\frac{1}{2}f_2^{(0,\gamma)}(x)\ln^2 \biggl( \frac{m_\mu^2}{m_e^2} \biggr)
+f_2^{(1,\gamma)}(x) \ln \biggl( \frac{m_\mu^2}{m_e^2} \biggr) + \ldots
\right ],
\label{photonfinal}
\ee
where the dots stand for ${\cal O}(\alpha^2)$ terms without the logarithmic
enhancement as well as higher order terms in the expansion of the 
electron energy
spectrum in the fine structure constant. The ${\cal O}(\alpha^{0})$ energy 
spectrum is given by $f_0(x) = x^2(3-2x)$. The ${\cal O}(\alpha)$ 
correction to it, $f_1(x)$, was calculated in Ref.~\cite{Kinoshita:1959ru}.
The coefficient of the double--log term is given by
\ba \label{f2LLg}
f_2^{(0,\gamma)}(x) &=& 4x^2(3-2x) \left [
\frac{1}{2}\ln^2x + \ln^2(1-x) - 2\ln x\ln(1-x)
- \Li_2(1-x) - \zeta_2 \right ]
\nonumber \\ 
&+& \biggl( \frac{10}{3} + 8x - 16x^2 + \frac{32}{3}x^3 \biggr)\ln(1-x)
+ \biggl( - \frac{5}{6} - 2x + 8x^2 - \frac{32}{3}x^3\biggr)\ln x
\nonumber \\
&+& \frac{11}{36} + \frac{17}{6}x + \frac{8}{3}x^2 - \frac{32}{9}x^3,
\ea
and is therefore in agreement with recent results in 
Ref.\cite{Arbuzov:2002pp}. 
The coefficient of the single--log enhanced term for pure photonic 
corrections, which is one of the new results in this Letter, reads
\ba \label{f2NLg}
f_2^{(1,\gamma)}(x) &=&
2x^2(3-2x)\biggl( - 2\Li_3(x) - 2\Sot{x}
+ 2\Li_2(x)\ln(1-x) + 2\Li_2(x)\ln x
\nonumber \\ 
&+& 5\ln x\ln^2(1-x) - 5\ln^2x\ln(1-x)
+ 2\ln^3x - 2\zeta_2\ln(1-x) - 2\zeta_2\ln x + 7\zeta_3 \biggr)
\nonumber \\ 
&+& \Li_2(x)\biggl( \frac{10}{3} + 14x - 40x^2 + \frac{92}{3}x^3 \biggr)
+ \ln x \ln(1-x) \biggl( \frac{25}{3} + 32x - 54x^2 + \frac{92}{3}x^3 \biggr)
\nonumber \\ 
&+& \ln^2(1-x) \biggl( - 12x - 4x^2 + 8x^3 \biggr)
+ \ln^2x \biggl( - \frac{25}{12} - 5x + 22x^2 - \frac{70}{3}x^3 \biggr)
\nonumber \\ 
&+& \ln(1-x) \biggl( - \frac{17}{3} - \frac{53}{3}x 
+ \frac{64}{3}x^2 - 12x^3 \biggr)
+ \ln x 
\biggl( - \frac{3}{4} + \frac{37}{6}x + \frac{4}{3}x^2 
+ \frac{44}{9}x^3 \biggr)
\nonumber \\ 
&+& \zeta_2 \biggl( - \frac{10}{3} - 2x + 35x^2 - \frac{98}{3}x^3 \biggr)
+ \frac{211}{216} - \frac{287}{12}x + \frac{83}{3}x^2 - \frac{559}{54}x^3,
\ea
where we used
\be
\Li_3(x) \equiv \int\limits_{0}^{x}\dd z\;\frac{\Li_2(z)}{z}\, ,\qquad
\Sot{x} \equiv \frac{1}{2}\int\limits_{0}^{x}
\dd z\;\frac{\ln^2(1-z)}{z}\, .
\ee

The correction  due to non--singlet electron--positron pairs, 
including  effects of the running coupling constant, reads
\be
\Delta^{(\NS)} =  
\biggl( \frac{\alpha}{2\pi} \biggr)^2 \left [ 
\frac{1}{3}f_2^{(0,\NS)}(x)\ln^2 \biggl( \frac{m_\mu^2}{m_e^2} \biggr)
+ f_2^{(1,\NS)}(x) \ln \biggl( \frac{m_\mu^2}{m_e^2} \biggr) + \ldots
\right ],
\label{nsingfinal}
\ee
with 
\ba
\label{f2LLNS} 
f_{2}^{(0,\NS)}(x) &=& 2x^2(3-2x)\ln\frac{1-x}{x}
+ \frac{5}{6} + 2x - 4x^2 + \frac{8}{3}x^3,
\\ \label{f2NLNS} 
f_{2}^{(1,\NS)}(x) &=& 
2x^2(3-2x)\biggl( - 2\Li_2(1-x) - \frac{2}{3}\ln x\ln(1-x)
+ \frac{2}{3}\ln^2(1-x) - \ln^2x - \frac{2}{3}\zeta_2 \biggr)
\nonumber \\
&+& \ln(1-x) \biggl( \frac{10}{9} - \frac{4}{3}x - \frac{46}{3}x^2 
+ 12x^3 \biggr)
+ \ln x \biggl(\frac{5}{9} + \frac{4}{3}x + 8x^2 - \frac{76}{9}x^3 \biggr)
\nonumber \\
&-& \frac{11}{6} - \frac{19}{3}x + \frac{100}{9}x^2 - \frac{64}{9}x^3.
\ea

Next, we derive the result for the singlet pair correction. Writing
\be
\Delta^{(\SS)} =  
\left( \frac{\alpha}{2\pi} \right)^2 \left[ 
\frac{1}{2}f_2^{(0,\SS)}(x)\ln^2 \biggl( \frac{m_\mu^2}{m_e^2} \biggr)
+f_2^{(1,\SS)}(x) \ln \biggl( \frac{m_\mu^2}{m_e^2} \biggr) + \ldots
\right],
\label{singeltfin}
\ee
we obtain
\ba
\label{f2LLS}
f_2^{(0,\SS)}(x) &=&
\frac{2}{3x} + \frac{17}{9}
+ 3x - \frac{14}{3}x^2 - \frac{8}{9}x^3
+ \biggl( \frac{5}{3} + 4x + 4x^2 \biggr)\ln x,
\\ \label{f2NLS}
f_2^{(1,\SS)}(x) &=&
\biggl[ \Li_2(1-x) + \ln x\ln(1-x) \biggr]
\biggl( \frac{5}{3} + 4x + 4x^2 \biggr)
+ \ln^2x \biggr( \frac{5}{2} + 6x + 4x^2 \biggr)
\nonumber \\
&+& \ln(1-x)\biggl( \frac{17}{9} + \frac{2}{3x} + 3x - \frac{14}{3}x^2 
- \frac{8}{9}x^3 \biggr)
+ \ln x\biggl( \frac{8}{9} + \frac{4}{3x} - \frac{5}{6}x 
- \frac{19}{3}x^2 \biggr)
\nonumber \\
&-& \frac{1}{3x} - \frac{67}{9} + \frac{43}{18}x + \frac{77}{18}x^2 
+ \frac{10}{9}x^3.
\ea

Finally, for the interference term one finds
\be
\Delta^{(\mathrm{int})} =  
\biggl( \frac{\alpha}{2\pi} \biggr)^2 \left [ 
f_2^{(1,{\mathrm{int}})}(x) \ln \biggl( \frac{m_\mu^2}{m_e^2} \biggr) + \ldots
\right ],
\label{intfinal}
\ee
and
\ba
\label{f2NLi}
f_2^{(1,{\mathrm{int}})}(x) &=&
2x^2(3-2x)\biggl( 2\Li_3(1-x) - 4\Sot{1-x} - 2\Li_2(1-x)\ln x \biggr)
\nonumber \\
&+& \Li_2(1-x)\biggl( \frac{5}{3} + 4x - 26x^2 + \frac{52}{3}x^3 \biggr)
+ \ln^2x \biggl(  - 9x^2 + \frac{26}{3}x^3 \biggr)
\nonumber \\
&+& \ln x \biggl(  - \frac{5}{3} - \frac{5}{3}x - \frac{28}{3}x^2 \biggr)
- \frac{62}{9} + \frac{41}{3}x - \frac{55}{3}x^2 + \frac{104}{9}x^3.
\ea

\section{Conclusions}

\begin{figure}[thbp]
\epsfig{file=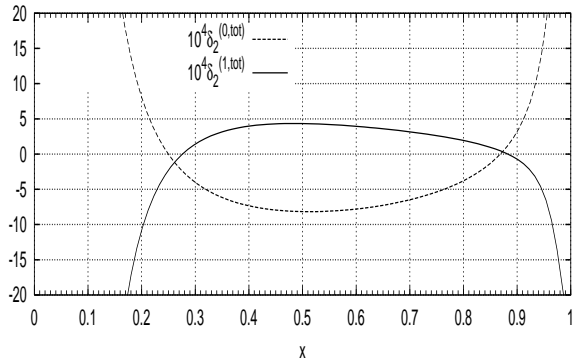,width=8cm,height=5cm} 
\caption{
Double and single logarithmic corrections as a function of $x$.}
\label{d2:fig}
\end{figure}

By applying techniques 
developed for  perturbative QCD to Quantum Electrodynamics,
we computed ${\cal O}(\alpha^2)$  correction to 
the electron energy spectrum in unpolarized muon decay, keeping
all the terms enhanced by the logarithms of the muon to 
electron mass ratio.
The double logarithmic corrections ${\cal O}(\alpha^2(\ln^2 m_\mu^2/m_e^2))$
are in agreement with recent 
result in Ref.~\cite{Arbuzov:2002pp}. 
The single logarithmic corrections
 ${\cal O}(\alpha^2 \ln(m_\mu^2/m_e^2))$,  the new result of this Letter,
are important to match  the precision requirements of the TWIST experiment. 
To illustrate significance of single logarithmic terms, 
in Fig.~\ref{d2:fig} we plot both double and single 
${\cal O}(\alpha^2)$ logarithmic corrections 
defined as
\ba \label{deltas}
&& \delta_2^{(0,\mathrm{tot})}(x) = 
\frac{1}{f_0(x)}\biggl(\frac{\alpha}{2\pi}\biggr)^2
\biggl[ \frac{1}{2}f_2^{(0,\gamma)}(x)
+ \frac{1}{3}f_2^{(0,\NS)}(x) + \frac{1}{2}f_2^{(0,\SS)}(x)
\biggr]\ln^2\biggl(\frac{m_\mu^2}{m_e^2}\biggr),
\nonumber \\
&& \delta_2^{(1,\mathrm{tot})}(x) = 
\frac{1}{f_0(x)}\biggl( \frac{\alpha}{2\pi} \biggr)^2
\biggl[ f_2^{(1,\gamma)}(x)
+ f_2^{(1,\NS)}(x) + f_2^{(1,\SS)}(x) + f_2^{(1,\mathrm{int})}(x)
\biggr]\ln\biggl(\frac{m_\mu^2}{m_e^2}\biggr),
\ea
as the function of the energy fraction carried away by electron in 
muon decay.

As follows from Fig.~\ref{d2:fig}, the 
$\order{\alpha^2\ln(m_\mu^2/m_e^2)}$ corrections computed in this Letter 
are required for the theoretical prediction with the precision $10^{-4}$, 
which is the benchmark precision for the TWIST experiment.
Moreover, within the acceptance region of the TWIST experiment, 
$0.3 \la x \la 0.98$, the magnitude 
of $\order{\alpha^2\ln(m_\mu/m_e)}$ corrections is in fact comparable 
to the magnitude of (naively the largest) $\order{\alpha^2\ln^2(m_\mu/m_e)}$ 
terms. 

The fact that sub--leading logarithmically enhanced corrections are
larger than the precision of the TWIST experiment, seems to indicate that 
full calculation of the ${\cal O}(\alpha^2)$ corrections to the 
electron energy spectrum is very desirable and that without such a calculation 
an intrinsic theory uncertainty can not be pushed below 
${\rm few} \times 10^{-4}$. 

Complete calculation of  ${\cal O}(\alpha^2)$  correction to the electron 
energy spectrum in muon decay is a very difficult task  and 
it remains to be seen if it can be done.
In spite of that, it is possible to extend the analysis of this 
Letter in a number of ways to further improve on the theory 
prediction. First of all, it is straightforward to compute the  electron 
energy spectrum in polarized muon decay, by using  
techniques described in this Letter. 
Furthermore, $\order{\alpha^3 \ln^3(m_\mu^2/m_e^2)}$ corrections
can be obtained from the DGLAP equation. 
Also,  resummation of the corrections that are singular in the limit 
$x \to 1$ can be performed and the influence of this resummation 
on  theoretical predictions for the spectrum can be explored.
These studies as well  as detailed discussion of 
the present theoretical uncertainty in the electron 
energy spectrum in polarized muon decay will be presented elsewhere.

\begin{acknowledgments}
This research was supported in part by the Natural
Sciences and Engineering Research Council of Canada and by the DOE
under grant number DE-AC03-76SF00515. We are grateful to A. Czarnecki
for attracting out attention to this problem and a number of useful 
conversation.
\end{acknowledgments}


\begin{thebibliography}{99}

\bibitem{Kuno:1999jp}
See e.g. Y.~Kuno and Y.~Okada,
Rev.\ Mod.\ Phys.\  {\bf 73}, 151 (2001), and references therein.

\bibitem{Rodning:2001js}
N.~L.~Rodning {\it et al.},
Nucl.\ Phys.\ Proc.\ Suppl.\  {\bf 98}, 247 (2001).

\bibitem{Quraan:2000vq}
M.~Quraan {\it et al.},
Nucl.\ Phys.\ A {\bf 663}, 903 (2000).

\bibitem{Kinoshita:2001pn}
T.~Kinoshita,
arXiv:hep-ph/0101197.

\bibitem{vanRitbergen:1998yd}
T.~van Ritbergen and R.~G.~Stuart,
Phys.\ Rev.\ Lett.\  {\bf 82}, 488 (1999).

\bibitem{Arbuzov:2002pp}
A.~Arbuzov, A.~Czarnecki, and A.~Gaponenko,
arXiv:hep-ph/0202102, to appear in Phys. Rev. D.

\bibitem{Ellis:qj}
R.~K.~Ellis, W.~J.~Stirling and B.~R.~Webber,
\textit{QCD And Collider Physics}
(Cambridge University Press, Cambridge, 1996)
and references therein.

\bibitem{Mele:1990cw}
B.~Mele and P.~Nason,
Nucl.\ Phys.\ B {\bf 361}, 626 (1991).

\bibitem{Cacciari:2001cw}
M.~Cacciari and S.~Catani,
Nucl.\ Phys.\ B {\bf 617}, 253 (2001).

\bibitem{Corcella:2001hz}
G.~Corcella and A.~D.~Mitov,
Nucl.\ Phys.\ B {\bf 623}, 247 (2002).

\bibitem{Kinoshita:1959ru}
T.~Kinoshita and A.~Sirlin,
Phys.\ Rev.\  {\bf 113}, 1652 (1959).

\bibitem{Arbuzov:2001ui}
A.~B.~Arbuzov,
Phys.\ Lett.\ B {\bf 524}, 99 (2002).

\bibitem{Curci:1980uw}
G.~Curci, W.~Furmanski and R.~Petronzio,
Nucl.\ Phys.\ B {\bf 175}, 27 (1980).

\bibitem{Floratos:1981hs}
E.~G.~Floratos, C.~Kounnas and R.~Lacaze,
Nucl.\ Phys.\ B {\bf 192}, 417 (1981).

\bibitem{Furmanski:1980cm}
W.~Furmanski and R.~Petronzio,
Phys.\ Lett.\ B {\bf 97}, 437 (1980).

\bibitem{Ellis:1996nn}
R.~K.~Ellis and W.~Vogelsang,
arXiv:hep-ph/9602356.

\bibitem{Berends:1987ab}
F.~A.~Berends, W.~L.~van Neerven and G.~J.~Burgers,
Nucl.\ Phys.\ B {\bf 297}, 429 (1988),
[Erratum-ibid.\ B {\bf 304}, 921 (1988)].


\end{thebibliography}
\end{document}